\documentclass[
  aps,  % which kind of journal: aps (default) or aip
  pra,
  superscriptaddress,  % Use superscriptaddress for affiliations
  raggedbottom,  % avoid large spaces between paragraphs
  floatfix,  % avoid problems with floats
  twocolumn,  % two column layout
  10pt,  % font size
  longbibliography,  % show the full titles in the bibliography, default true
  noeprint,  % remove the eprint information from the references
  nofootinbib,  % footnotes not in bibliography
]{revtex4-2}

\usepackage[T1]{fontenc}  % use modern output font encodings
\usepackage[utf8]{inputenc}  % use UTF-8 input encoding

\usepackage[USenglish]{babel}
\usepackage{amsmath}
\usepackage{amssymb}
\usepackage{graphicx}
\usepackage[dvipsnames]{xcolor}
\usepackage[pdfusetitle]{hyperref}
\usepackage{cleveref}
\usepackage{dsfont}
\usepackage{float}
\usepackage[normalem]{ulem}
\usepackage{siunitx}
\usepackage{MnSymbol}
\usepackage{bbm}
\usepackage{orcidlink}
\usepackage[USenglish]{isodate}
\usepackage{microtype}

\hypersetup{
  pdfsubject={Atomic Physics},  % suggestion: stay consistent with the arxiv category
  pdfauthor={Johannes Mögerle, Frederic Hummel, Alicia Keil, Tangi Legrand, Eduard J. Braun, Henri Menke, Jonathan King, Beatriz Olmos, Sebastian Hofferberth, Hans Peter Büchler, Sebastian Weber},
  colorlinks=true,  % false: boxed links; true: colored links
  linkcolor={red!60!black},  % color of internal links
  citecolor={blue!80!black},  % color of links to bibliography
  urlcolor={blue!95!black},  % color of external links
}

\newcommand{\abs}[1]{\left|{#1}\right|}  % e.g. \abs{x} -> |x|
\newcommand{\hc}{\text{h.c.}}  % e.g. H + \hc -> H + h.c.
\newcommand{\ket}[1]{\left|{#1}\right\rangle}  % e.g. \ket{0} -> |0>
\newcommand{\bra}[1]{\left\langle{#1}\right|}  % e.g. \bra{0} -> <0|
\newcommand{\braopket}[3]{\left\langle{#1}\middle|{#2}\middle|{#3}\right\rangle}  % e.g. \braopket{0}{1}{2} -> <0|1|2>
\newcommand{\ketbra}[2]{\ket{#1} \! \bra{#2}}  % e.g. \ketbra{0}{1} -> |0><1|

\newcommand{\state}[1]{\lvert{#1}\rangle}
\renewcommand{\vec}[1]{\mathbf{#1}}
\newcommand{\vechat}[1]{\hat{\mathbf{#1}}}
\renewcommand{\tensor}[1]{\mathbf{#1}}

\newcommand{\sr}{\ensuremath{s}}
\newcommand{\lr}{\ensuremath{l}}
\newcommand{\jr}{\ensuremath{j}}
\newcommand{\ic}{\ensuremath{I}}
\renewcommand{\sc}{\ensuremath{S_c}}
\newcommand{\lc}{\ensuremath{L_c}}
\newcommand{\jc}{\ensuremath{J_c}}
\newcommand{\fc}{\ensuremath{F_c}}
\newcommand{\stot}{\ensuremath{S}}
\newcommand{\ltot}{\ensuremath{L}}
\newcommand{\jtot}{\ensuremath{J}}
\newcommand{\ftot}{\ensuremath{F}}

\begin{document}
% % % % % % % % % % PREAMBLE % % % % % % % % % %
\title{The PairInteraction Toolkit for Modeling Rydberg Physics in Alkali and Alkaline-Earth-Like Atoms}

\author{Johannes Mögerle\,\orcidlink{0000-0002-7761-9130}}
\email{johannes.moegerle@itp3.uni-stuttgart.de}
\affiliation{Institute for Theoretical Physics III and Center for Integrated Quantum Science and Technology, University of Stuttgart, Pfaffenwaldring 57, 70569 Stuttgart, Germany}

\author{Frederic Hummel\,\orcidlink{0000-0003-0180-6615}}
\affiliation{Atom Computing, Inc., 1901 Fourth St, Ste 200, Berkeley, California, 94710, USA}

\author{Alicia Keil\,\orcidlink{0009-0003-1129-8543}}
\affiliation{Institut für Theoretische Physik, Universität Tübingen, Auf der Morgenstelle 14, 72076 Tübingen, Germany}

\author{Tangi Legrand\,\orcidlink{0009-0001-1394-5738}}
\affiliation{Institute of Applied Physics, University of Bonn, Wegelerstraße 8, 53115 Bonn, Germany}

\author{Eduard J. Braun\,\orcidlink{0009-0009-5061-3036}}
\affiliation{Physikalisches Institut, Universität Heidelberg, Im Neuenheimer Feld 226, 69120 Heidelberg, Germany}

\author{Henri Menke\,\orcidlink{0000-0001-5245-5400}}
\affiliation{Max Planck Computing and Data Facility, Gießenbachstraße 2, 85748 Garching b. München, Germany}

\author{Jonathan King}
\affiliation{Atom Computing, Inc., 1901 Fourth St, Ste 200, Berkeley, California, 94710, USA}

\author{Beatriz Olmos\,\orcidlink{0000-0002-1140-2641}}
\affiliation{Institut für Theoretische Physik, Universität Tübingen, Auf der Morgenstelle 14, 72076 Tübingen, Germany}

\author{Sebastian Hofferberth\,\orcidlink{0000-0003-0309-9715}}
\affiliation{Institute of Applied Physics, University of Bonn, Wegelerstraße 8, 53115 Bonn, Germany}

\author{Hans Peter Büchler\,\orcidlink{0000-0002-7233-6828}}
\affiliation{Institute for Theoretical Physics III and Center for Integrated Quantum Science and Technology, University of Stuttgart, Pfaffenwaldring 57, 70569 Stuttgart, Germany}

\author{Sebastian Weber\,\orcidlink{0000-0001-9763-9131}}
\affiliation{Institute for Theoretical Physics III and Center for Integrated Quantum Science and Technology, University of Stuttgart, Pfaffenwaldring 57, 70569 Stuttgart, Germany}

\date{\printdate{2026-05-06}}


% % % % % % % % % % ABSTRACT % % % % % % % % % %
\begin{abstract}
  Rydberg atoms provide a powerful platform for exploring strongly interacting quantum systems, both in free space and in structured electromagnetic environments, with growing applications in quantum technology.
  Accurately modeling their single-atom properties and mutual interactions is essential for interpreting experiments and designing new architectures.
  We present a unified theoretical framework for Rydberg atoms and their interactions based on multi-channel quantum defect theory (MQDT) and static electromagnetic Green's tensors.
  MQDT provides a precise description of Rydberg states of divalent atoms such as strontium and ytterbium, while the Green's tensor formalism provides a general and flexible approach for calculating interactions between two Rydberg atoms in arbitrary geometries, including modifications induced by nearby surfaces.
  We implement this framework in an updated version of the open-source toolkit \textit{PairInteraction} [Weber et al., J.~Phys.~B~50 (2017)].
  The implementation leverages high-performance libraries and achieves speedups of one order of magnitude for pair-potential calculations compared to prior software.
  We demonstrate the capabilities of the toolkit through example applications to divalent atoms and show excellent agreement with experimental data for an exemplary Stark map of $^{174}$Yb.
  The modular software architecture enables the community to extend it further.
\end{abstract}

% % % % % % % % % % MAKETITLE % % % % % % % % % %
\maketitle

% % % % % % % % % % MAINPART % % % % % % % % % %
\section{Introduction\label{sec:introduction}}

Atoms excited to Rydberg states~\cite{gallagher_rydberg_1994} exhibit strong, long-range interactions that are of fundamental interest and underpin a wide range of applications.
In neutral-atom quantum simulation and computing, these interactions enable entanglement via direct Rydberg coupling~\cite{jaksch_fast_2000, wilk_entanglement_2010, levine_parallel_2019, madjarov_highfidelity_2020,graham_multiqubit_2022, anand_dualspecies_2024, cao_multiqubit_2024,bluvstein_logical_2024} or Rydberg dressing~\cite{jau_entangling_2016, zeiher_manybody_2016, schine_longlived_2022, weckesser_realization_2025}.
The strong field sensitivity of Rydberg states enables precision metrology~\cite{sedlacek_microwave_2012, holloway_subwavelength_2014}, where interaction-induced shifts can further enhance sensitivity~\cite{ding_enhanced_2022, wang_highprecision_2025}.
Moreover, the coherent mapping of Rydberg interactions onto quantized optical fields enables strong optical nonlinearities, with applications in optical quantum information processing~\cite{gorshkov_photonphoton_2011, peyronel_quantum_2012, firstenberg_attractive_2013,firstenberg_nonlinear_2016, drori_quantum_2023}.
In parallel, Rydberg atoms have been studied in structured electromagnetic environments, including optical cavities~\cite{sheng_intracavity_2017, suarez_superradiance_2022}, near surfaces~\cite{thiele_manipulating_2014, ocola_control_2024,block_van_2017}, and in hybrid quantum systems~\cite{morgan_coupling_2020, kaiser_cavitydriven_2022}, which require corresponding theoretical descriptions.
In recent years, divalent atoms, also referred to as alkaline-earth-like atoms, have attracted significant interest due to their rich internal structure~\cite{dunning_recent_2016, cooper_alkalineearth_2018, jenkins_ytterbium_2022, ma_highfidelity_2023,holzl_longlived_2024,muniz_highfidelity_2025}.
They offer a wide range of spin-encoding possibilities~\cite{barnes_assembly_2022,chen_analyzing_2022,unnikrishnan_coherent_2024,pucher_finestructure_2024, jia_architecture_2024, ammenwerth_realization_2025}, while their core electron can be exploited for coherent light shifts~\cite{madjarov_highfidelity_2020, burgers_controlling_2022, pham_coherent_2022}, state readout~\cite{lochead_numberresolved_2013, madjarov_highfidelity_2020}, and magic trapping~\cite{mukherjee_manybody_2011, wilson_trapping_2022}.
At the same time, their Rydberg spectra are significantly more intricate than in alkali atoms: core-excited states can lie close in energy to the relevant Rydberg series and perturb them, particularly in systems with hyperfine-split cores~\cite{robicheaux_theory_2018}.
As a result, a multi-channel quantum defect theory (MQDT) description is required~\cite{seaton_quantum_1966,fano_quantum_1970,vaillant_multichannel_2014,robicheaux_theory_2018,hummel_engineering_2024} rather than the commonly used single-channel treatment (SQDT)~\cite{deiglmayr_longrange_2016,weber_calculation_2017}.

For interpreting experiments and designing new architectures, calculations beyond perturbative treatments~\cite{comparat_dipole_2010, vaillant_longrange_2012, robicheaux_calculations_2019} are required.
However, established numerical tools for non-perturbative calculations, such as ARC~\cite{robertson_arc_2021} and the previous version of PairInteraction~\cite{weber_calculation_2017}, have so far been restricted to SQDT models.
Recently, specialized codes for specific MQDT use cases have emerged, such as for fitting MQDT parameters~\cite{potvliege_mqdtfit_2024} and for ytterbium calculations~\cite{peper_spectroscopy_2025}.
To provide a user-friendly and flexible toolkit that also supports MQDT models, we have released PairInteraction v2 as open-source software: \url{www.pairinteraction.org}.
Furthermore, support for the Green's tensor formalism for calculating Rydberg interactions in structured electromagnetic environments has been enhanced.
The modular architecture makes the software extensible and provides a high-performance backend that accelerates calculations by one order of magnitude over the previous version.
Moreover, the software allows users to extract effective Hamiltonians, for example, to model many-body Rydberg quantum simulators beyond the description in terms of dispersion coefficients.

In this paper, we discuss MQDT and the Green's tensor formalism for modeling interacting Rydberg atoms, and their implementation in PairInteraction.
With respect to MQDT, we propose a scheme for unambiguously labeling Rydberg states by averaged angular quantum numbers.
Regarding the Green's tensor formalism, we show the importance of couplings due to self-interaction.
After discussing the theory, we present the new architecture of PairInteraction, highlighting design decisions to improve performance and usability.
We conclude by discussing two exemplary use cases of the software.
First, we study the single- and two-atom Rydberg physics of ytterbium in external fields, demonstrating excellent agreement with experimentally measured Stark shifts and extracting effective spin-1/2 and spin-1 model Hamiltonians.
Second, we calculate Rydberg interactions near a surface, analyzing the resulting state mixing and modifications to the van der Waals coefficients.

\section{Setup\label{sec:setup}}
\begin{figure}[t]
  \centering
  \includegraphics[width=\columnwidth]{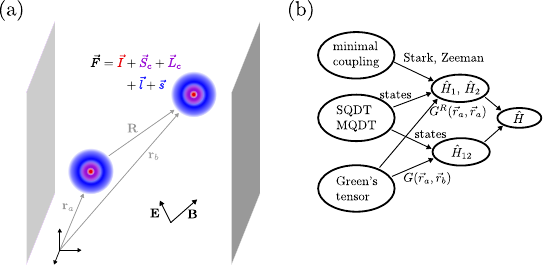}
  \caption{
    \textbf{Considered system.} (a) Two interacting Rydberg atoms, either in free space or in a structured electromagnetic environment (the figure depicts a cavity).
    The atoms can be alkali atoms or divalent atoms.
    Static electric and magnetic fields can be applied in arbitrary directions.
    (b) Overview of the approach used to construct the system's Hamiltonian, starting with the formalisms applied to describe the physics.
  }\label{fig:setup}
\end{figure}

We consider a system of two Rydberg atoms, either in free space or in a structured electromagnetic environment, such as a cavity or near a conducting plate.
Static electric or magnetic fields in arbitrary directions may be present, see \cref{fig:setup}(a).
We restrict ourselves to atoms with a single valence electron excited to a Rydberg state.
The atoms can be of any species for which sufficient spectroscopic data is available in the literature to model them via quantum defect theory.
This includes alkali atoms and well-characterized divalent species such as strontium and ytterbium.
The two atoms may also belong to different species.

In this work, we focus on the electronic Hamiltonian and assume the validity of the Born-Oppenheimer approximation~\cite{born_zur_1927}.
The total Hamiltonian of the system is given by the sum of single-atom contributions and two-body interactions (see \cref{fig:setup}(b)),
\begin{equation}
  \hat{H} =
  \hat{H}_{1} \otimes \mathbbm{1} + \mathbbm{1} \otimes \hat{H}_{2} + \hat{H}_{12}
  \,.
\end{equation}
The single-atom Hamiltonian is written as
$\hat{H}_{\alpha} = \hat{H}_{0,\alpha} + \hat{H}_{\mathrm{fields},\alpha} + \hat{H}_{\mathrm{si},\alpha}$ with $\alpha=1,2$, where $\hat{H}_{0,\alpha}$ describes the field-free Hamiltonian containing the energies of the unperturbed atomic levels, $\hat{H}_{\mathrm{fields},\alpha}$ the coupling to static external fields, and $\hat{H}_{\mathrm{si},\alpha}$ the self-interaction induced by the structured electromagnetic environment.
The interaction between the two atoms is given by $\hat{H}_{12}$.

The energies of the unperturbed atomic levels always include the fine-structure splitting.
In the case of divalent atoms with non-zero nuclear spin, the hyperfine splitting must be taken into account as well, because the hyperfine structure of the ionic core is on relevant energy scales.
For atoms with a single valence electron, the calculation of the energies and corresponding states via single-channel quantum defect theory (SQDT) is a standard technique~\cite{seaton_quantum_1958,seaton_quantum_1983}.
If ionic-core excitations are relevant, as is commonly the case for divalent atoms, multi-channel quantum defect theory (MQDT) is required~\cite{seaton_quantum_1966,fano_quantum_1970}.
Although we neglect ionization channels, they could, in principle, be included via complex absorbing potentials~\cite{riss_calculation_1993,muga_complex_2004}.

The interaction with static external electric and magnetic fields can be treated within the minimal coupling framework, leading to Stark and Zeeman Hamiltonians, including diamagnetic terms~\cite{steck_quantum_2007,milonni_quantum_1994}.
This treatment is not restricted to homogeneous fields, but is also applicable in the case of inhomogeneous fields, which can be used to model ion-induced Stark shifts.

The interaction between atoms as well as the self-interaction, e.g., when the atoms are near conducting plates, is treated in the Green's tensor formalism~\cite{lehmberg_radiation_1970,jones_modified_2018,dung_resonant_2002,buhmann_dispersion_2012,asenjo-garcia_atomlight_2017,fuchs_casimirpolder_2017,ribeiro_casimirpolder_2015,block_van_2017,block_casimirpolderinduced_2019}.
This formalism allows for arbitrary geometrical arrangements and can account for the effects of structured electromagnetic environments.
Here, we assume that atom-atom and atom-surface distances are small compared with the wavelengths of relevant transitions so that we can use the static Green's tensor~\cite{block_casimirpolderinduced_2019, block_van_2017}.
For simple conducting geometries, this static Green's tensor treatment is equivalent to the method of image charges~\cite{buhmann_dispersion_2012}.
Moreover, the distances must be larger than the LeRoy radius~\cite{leroy_longrange_1974} so that wave functions do not overlap, allowing us to neglect exchange interactions and truncate the interaction order.
Both assumptions are typically satisfied in experiments with Rydberg-atom arrays and provide a reasonable description of many cold-atomic-cloud experiments.

In the following sections, we focus on MQDT and the Green's tensor formalism, as these topics have both become increasingly relevant for Rydberg platforms in recent years and are not yet as comprehensively summarized in the literature.
However, our open-source toolkit PairInteraction also incorporates all other physical effects described above~\cite{weber_calculation_2017}.

\section{Multi-Channel Quantum Defect Theory\label{sec:mqdt}}

In this section, we discuss how Rydberg states of atoms with multiple valence electrons can be described by multi-channel quantum defect theory (MQDT)~\cite{seaton_quantum_1966,fano_quantum_1970}.
Quantum defect theory applies to systems in which one electron is excited to a Rydberg state, while the nucleus and the remaining electrons form an ionic core.
The following summary of MQDT combines discussions from the primer by S. Ross~\cite{ross_mqdt_1991}, work by C. Greene et al.~\cite{aymar_multichannel_1996}, and publications by F. Robicheaux et al.~\cite{robicheaux_theory_2018}, C. L. Vaillant et al.~\cite{vaillant_multichannel_2014}, and M. Peper et al.~\cite{peper_spectroscopy_2025}.

\subsection{Wave Functions in FJ-coupled channels\label{sec:radial}}

In quantum defect theory, a channel is specified by a set of angular-momentum quantum numbers of the Rydberg electron and the ionic core, in a specific coupling scheme, together with the residual core configuration.
If the Rydberg electron is far away from the ionic core, a description in the FJ-coupling scheme is most appropriate.
In this representation, a channel $i$, also referred to as collision or ionization channel, is described by the Rydberg electron quantum numbers $\ket{\sr, \lr, \jr, m_{\jr}}$ and the ionic core quantum numbers $\ket{\ic, \sc, \lc, \jc, \fc, m_{\fc}}$.
Here, $\sr$, $\lr$, $\jr$ denote the spin, orbital, and total angular momentum of the single Rydberg electron,
while $\sc$, $\lc$, $\jc$ denote the total spin, total orbital, and total angular momentum of all core electrons.
The quantum numbers $\ic$ and $\fc$ are the nuclear spin and the hyperfine angular momentum of the ionic core.
All remaining properties of the core, such as the radial wave functions of the core electrons and their remaining angular quantum numbers, are included in the residual core configuration $\ket{c}$.
The total angular wave function is obtained by coupling $\fc$ and $\jr$ to the total angular momentum $\ftot$, whose projection along the quantization axis yields $m_{\ftot}$.
The resulting FJ-coupled channel is given by
\begin{equation}
  \ket{\phi_i} = \ket{\ic, \sc, \lc, \jc, \fc; \sr, \lr, \jr; \ftot, m_{\ftot}} \ket{c}
  \,.
\end{equation}
The product of $\ket{\phi_i}$ and an associated radial wave function $\psi_i(r)$ defines a so-called channel function.
We will see later on that Rydberg states can, in general, be described by superpositions of such channel functions.

\begin{figure}[t]
  \centering
  \includegraphics{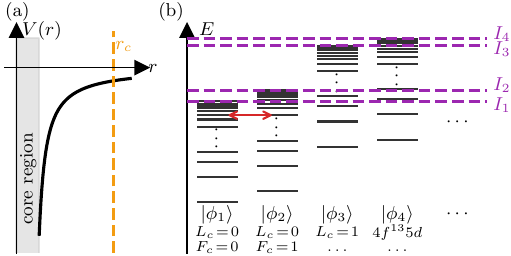}
  \caption{\textbf{MQDT concepts.}
    (a) The calculation of Rydberg energies and channel coefficients with MQDT typically assumes that, beyond a radius $r_c$, the potential of the ionic core is a pure Coulomb potential.
    As an approximation, Coulomb wave functions can remain useful at shorter distances up to a core region where irregular Coulomb functions diverge, and multi-electron effects dominate.
    (b) Schematic sketch of the FJ-coupled channels relevant to the $S_{1/2}$ Rydberg series of $^{171}$Yb.
    In this example, $\ket{\phi_1}$ and $\ket{\phi_2}$ are hyperfine split ground states of the ionic core.
    Due to the small hyperfine splitting, resonances appear (symbolized by a red arrow) which lead to almost perfect LS-coupled states (see also \cref{fig:mqdt-exp-qn}).
    The channels  $\ket{\phi_3}, \ket{\phi_4}, \dots$ are core excited states and their Rydberg series contribute to the bound state structure with only a few perturber states.
  }\label{fig:qdt-theory}
\end{figure}

The central idea behind quantum defect theory is to divide the real space around the ionic core into a complicated inner region $r \leq r_c$ and an outer region $r > r_c$, where $r$ denotes the distance from the ionic core.
In the outer region, the system can be treated as a single electron in a Coulomb potential, see \cref{fig:qdt-theory}(a).
Thus, the radial wave function in the outer region can be calculated for each FJ-coupled channel separately.

For $r > r_c$, the Schrödinger equation for the rescaled radial wave function $u_i(r) = r \psi_i(r)$ reads in reduced-mass atomic units, i.e., atomic units in which the electron's \textit{reduced} mass is set to unity,
\begin{align}
  \left( -\frac{1}{2}\frac{\mathrm{d}^2}{\mathrm{d}r^2} + \frac{l_i(l_i+1)}{2 r^2} + V(r) \right) u_i(r)  =  \mathcal{E}_i u_i(r)
  \,.
  \label{eqn:radial_seq}
\end{align}
The quantum number $l_i$ is the Rydberg electron orbital angular momentum of the $i$-th channel, and $V(r) = -\frac{1}{r}$ is the Coulomb potential, which resembles the potential of the ionic core for large distances.
In principle, one could also incorporate deviations from the Coulomb potential in the Schrödinger equation~\cite{fu_multiscale_2016}.
Note, however, that in literature, MQDT parameters are most often provided for a pure Coulomb potential.

The radial Schrödinger equation can be solved analytically.
For each FJ-coupled channel $i$, it has two linearly independent solutions with the same bound-state energy $\mathcal{E}_i$: the regular Coulomb function $f_{l_i}(\nu_i, r)$ and the Coulomb function $g_{l_i}(\nu_i, r)$ that is irregular at $r=0$, where we adopt the convention used in~\cite{seaton_quantum_1983,aymar_multichannel_1996}.
These functions are parameterized by the effective principal quantum number $\nu_i$, defined via
\begin{align}
  \mathcal{E}_i = -\frac{1}{2 \nu_i^2}
  \,.
  \label{eqn:eryd}
\end{align}
Then, the energy $E$ of the total atomic wave function is the sum of the energy of the Rydberg electron $\mathcal{E}_i$ and the energy of the ionic core.
Taking the atomic ground state energy as the zero of energy, the ionic core energy is given by the ionization threshold $I_i$ of the FJ-coupled channel such that $E = \mathcal{E}_i + I_i$.

\subsection{MQDT Wave Functions}\label{sec:mqdt-theory}

The ionic core can induce couplings between different FJ-coupled channels with the same parity and total angular momentum $\ftot$ (this follows from inversion symmetry and isotropy of the interaction between the Rydberg electron and the ionic core).
Couplings between channels are typical for divalent atoms.
For these atoms, the remaining valence electron allows for core excitations, in addition to one electron being excited to a Rydberg state.
To construct a total wave function that is an eigenfunction in this setting, we must build a linear combination of FJ-coupled channel functions.
The channel functions must belong to the same total energy $E$, i.e., the effective principal quantum numbers $\nu_i$ of the channel functions are given by $E = I_i - \frac{1}{2\nu_i^{2}}$.
We write the total wave function as
\begin{align}
  \Psi(\vec{r}) = \frac{1}{r} \sum_i \Bigl( a_i f_{l_i}(\nu_i, r) +
  b_i g_{l_i}(\nu_i, r)  \Bigr) \ket{\phi_i}
  \,.
  \label{eqn:mqdt-Psi}
\end{align}
In the context of MQDT, it has proven useful to reparameterize the coefficient vector $\vec{b}$ as
\begin{align}
  \vec{b} = - \boldsymbol{K} \vec{a}
  \,,
  \label{eqn:mqdt-b}
\end{align}
where the real and symmetric matrix $\boldsymbol{K}$ is known as the short-range reaction matrix.
The K-matrix incorporates the boundary condition that, at $r=r_c$, the wave function must continuously connect to the wave function of the inner region.
It captures deviations of the potential of the ionic core from a pure Coulomb potential, and complicated multi-electron physics inside the ionic core region.
The K-matrix is only weakly energy-dependent~\cite{lee_spectroscopy_1973,burke_multichannel_1998}, so that the same parameterization of the K-matrix works for many different Rydberg states.
Thus, while ab-initio calculations of the K-matrix would require demanding numerical simulations~\cite{child_abinitio_2011}, it can be readily determined from fits to experimental data~\cite{robicheaux_calculations_2019,peper_spectroscopy_2025,kuroda_microwave_2025}.

To incorporate the boundary condition that the wave function must vanish at $r \rightarrow \infty$ for bound states, one uses the asymptotic behavior of the Coulomb functions $f_{l} \rightarrow C(r)\;\sin(\pi(\nu_i-l))\;e^{r/\nu_i}$ and $g_{l} \rightarrow -C(r)\;\cos(\pi(\nu_i-l))\;e^{r/\nu_i}$~\cite{ross_mqdt_1991}.
The function $C(r)$ is not of importance, except for it going to zero more slowly than the exponential factor blowing up.
Thus, for the wave function to vanish at $r \rightarrow \infty$, it must be satisfied that
\begin{equation}
  \bigl(
  \tan(\pi \boldsymbol{\nu}) + \boldsymbol{K}
  \bigr)
  \vec{a}
  = \vec{0}
  \,,
  \label{eqn:mqdt-bc}
\end{equation}
where $\boldsymbol{\nu}$ is defined as diagonal matrix with the effective principal quantum numbers on its diagonal, $(\boldsymbol{\nu})_{ii} = \nu_i$.
A non-trivial solution exists only if the determinant condition
\begin{equation}
  \det\bigl(\tan(\pi \boldsymbol{\nu})+\boldsymbol{K}\bigr) = 0
  \label{eqn:det-cond-mqdt}
\end{equation}
is fulfilled.

To find an efficient representation of the K-matrix, we can first diagonalize it
\begin{equation}
  \boldsymbol{K} = \boldsymbol{Q}\; \boldsymbol{\tilde{K}} \;\boldsymbol{Q}^T
  \,,
\end{equation}
with the orthogonal matrix $\boldsymbol{Q}$.
The entries of the diagonal matrix $\boldsymbol{\tilde{K}}$ are sometimes written as $\tilde{K}_{\alpha\alpha} = \tan(\pi \tilde{\mu}_\alpha)$ with eigen quantum defects $\tilde{\mu}_\alpha$.

In the following, we show how an analytic ansatz for $\boldsymbol{Q}$ can be developed to reduce the number of parameters required to describe the K-matrix.
A useful approach is ``multi-channel quantum defect theory with a frame transformation'' (MQDT-FT)~\cite{burke_multichannel_1998,robicheaux_theory_2018}.
This technique relies on the insight that the K-matrix primarily accounts for effects at short distances where the relevant energy scales are large compared to the differences between the ionization thresholds $I_i$, which permits treating them as degenerate.
This allows one to define new channels that are only weakly mixed.
These are typically channels in which the Rydberg electron and the core are LS-coupled.
The frame transformation $\boldsymbol{Q}_{FJ-LS}$ from the LS-coupled channels to the FJ-coupled channels can be derived analytically~\cite{edmonds_angular_1957,robicheaux_theory_2018}.
Using this, a useful ansatz for $\boldsymbol{Q}$ is
\begin{align}
  \boldsymbol{Q} = \boldsymbol{Q}_{FJ-LS} \prod \boldsymbol{R}_{\alpha,\beta}
  \,,
  \label{eqn:frame-trafo-and-rotations}
\end{align}
where $\boldsymbol{R}_{\alpha,\beta}$ are rotation matrices between the LS-coupled channels $\alpha$ and $\beta$ about some specified angle, allowing for additional couplings (note that the order of the rotations matters).

The rotation angles and eigen quantum defects can be approximated by an expansion in energy.
The expansion coefficients are tabulated in literature, especially for atomic species that are of interest for applications in quantum technology, such as strontium~\cite{vaillant_multichannel_2014, robicheaux_calculations_2019} or ytterbium~\cite{lehec_laser_2018, peper_spectroscopy_2025,kuroda_microwave_2025}.

To get energies and wave functions of Rydberg states, we apply the above relations in reverse order:
Starting from the tabulated rotation angles and eigen quantum defects, one calculates the K-matrix in the collision-channel representation.
Then, one solves the determinant condition \cref{eqn:det-cond-mqdt} together with the energy condition $E = I_i - \frac{1}{2\nu_i^{2}}$, to obtain values for the energy $E$ and the effective principal quantum numbers $\nu_i$, using tabulated ionization thresholds $I_i$.
We get the coefficient vector $\vec{a}$ as the null vector from \cref{eqn:mqdt-bc}.
Using \cref{eqn:mqdt-bc} and \cref{eqn:mqdt-Psi,eqn:mqdt-b} the wave function for $r > r_c$ is then given by
\begin{align}
  \Psi(\vec{r})
  = \sum_i A_i \Biggl[ \frac{N_i}{r} \biggl( f_{l_i}(\nu_i, r) + \tan(\pi \nu_i) g_{l_i}(\nu_i, r) \biggr) \Biggr] \ket{\phi_i}
  \,,
  \label{eqn:mqdt-Psi-final}
\end{align}
where $N_i$ is chosen such that the radial part in the square brackets is normalized.
The channel coefficients are given by $A_i = a_i / N_i$, together with the normalization condition $\sum_i |A_i|^2 = 1$.
The normalization condition is required because we know the vector $\vec{a}$ only up to its norm.

Note that for obtaining energies and the coefficient vector $\vec{a}$ with MQDT, we assumed that the potential of the ionic core is a pure Coulomb potential beyond some arbitrary large radius $r_c$.
To get the wave function up to the core region, one typically assumes that the Coulomb potential is a good approximation also at shorter distances so that \cref{eqn:mqdt-Psi-final} can be used also there.
Under this assumption, the channel normalization factors are
$N_i = \frac{\nu_i^{3/2}}{\cos(\pi \nu_i)}$.
Corrections can be calculated from the energy dependence of the K-matrix~\cite{lee_spectroscopy_1973,seaton_quantum_1983}.

\subsection{Labeling\label{sec:mqdt-labeling}}

\begin{figure}[t]
  \centering
  \includegraphics{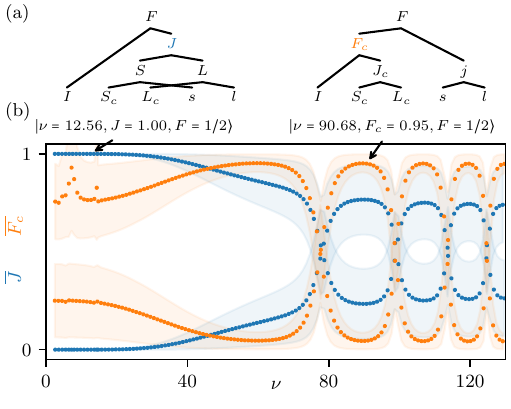}
  \caption{\textbf{Labeling the angular character of the $S_{1/2}$ Rydberg series of $^{171}$Yb.} (a) Sketch of the LS-coupling scheme (left) and FJ-coupling scheme (right).
    For the considered series, $\jtot$ and $\fc$ are non-trivial quantum numbers.
    (b) Average value of these quantum numbers (dotted) and their standard deviation (shaded region) as a function of the effective principal quantum number $\nu$.
    For low $\nu$, the states are almost perfectly LS-coupled, and therefore $\jtot$ is a good quantum number.
    For high $\nu$, most of the states are described well in the FJ-coupling scheme, and $\fc$ is a good quantum number.
    For some high $\nu$ states, neither $\jtot$ nor $\fc$ is a good quantum number.
    This Rydberg series is a simple example, where only two channels and some perturber states from other channels contribute, see \cref{fig:qdt-theory}(b).
  }\label{fig:mqdt-exp-qn}
\end{figure}

A Rydberg state obtained with MQDT is a superposition of different channel functions, and thus usually does not have well-defined quantum numbers apart from $\ftot$ and parity.

Instead of the principal quantum number $n$, one can use an effective principal quantum number
\begin{align}
  \nu = \frac{1}{\sqrt{2(I_\text{ref}-E)}}
\end{align}
to label a state, which is calculated from the energy $E$ of the state.
Here, we define $I_\text{ref}$ as the lowest ionization threshold of all existing channels of an atomic species.

We propose characterizing the angular part of a state by the average value and standard deviation of the angular quantum numbers $f \in \{\lr, \jr, \lc, \jc, \fc, \stot, \ltot, \jtot\}$:
\begin{align}
  \overline{f} &= \sum_{i} \abs{A_i}^2 f_i
  \,, \nonumber \\
  \Delta f &= \sqrt{\sum_{i} \abs{A_i}^2 (f_i - \overline{f})^2}
  \,,
\end{align}
where $f_i$ is the value of the quantum number $f$ of the channel $i$, and $A_i$ is its channel coefficient after transforming the total wave function in \cref{eqn:mqdt-Psi-final} into a coupling scheme where $f$ is a good quantum number of the channel wave functions.
In \cref{fig:mqdt-exp-qn}, we apply this concept to the $S_{1/2}$ series of $^{171}$Yb.
For this series, some states are better described by $\jtot$ (i.e., in the LS-coupling scheme), while others are better described by $\fc$ (i.e., in the FJ-coupling scheme).
Thus, to find meaningful labels for the angular part of a Rydberg state, we suggest identifying the coupling scheme for which the variance of the angular quantum numbers is smallest.
For some states, neither $\jtot$ nor $\fc$ is a good quantum number, which corresponds to states that are in a superposition of different channels in both the LS-coupling and the FJ-coupling scheme, because the angular part cannot be transformed independently of the radial part.
Our proposal is analogous to the labeling of molecular states, where one assigns term symbols using the coupling scheme that yields the most nearly ``good'' quantum numbers~\cite{strauss_hyperfine_2010}.

\subsection{Matrix Elements}

To calculate the interaction of Rydberg atoms with external fields and with each other, one must calculate matrix elements of electric multipole and angular-momentum operators~\cite{weber_calculation_2017}.

For electric multipole operators $\hat{d}_{\kappa q}$, one can ignore contributions from the core electrons and only consider the multipole moment of the Rydberg electron,
since the radial expectation values between Rydberg states scale as $\braopket{\Psi(\vec{r})}{\hat{r}^\kappa}{\Psi'(\vec{r})} \sim n^{2 \kappa}$,
so that the contributions of the tightly bound core electrons are negligible in comparison.
However, if one is interested in matrix elements between Rydberg states and low-lying states, e.g., for calculating Rydberg lifetimes,
the contribution of the second valence electron in divalent atoms can become significant~\cite{vaillant_multichannel_2014}.
To include this contribution, one also has to consider the angular and radial parts of the inner electrons.

The multipole matrix element between two states $\ket{\Psi}$ and $\ket{\Psi'}$, ignoring core contributions, is given by
\begin{align}
  \braopket{\Psi}{\hat{d}_{\kappa q}}{\Psi'} = \sum_{i,j} A_i^* A'_j \braopket{\phi_i}{\hat{Y}_{\kappa q}}{\phi'_j} \int_0^\infty dr \, r^{2 + \kappa} \psi_i(r) \psi'_j(r)
  \,,
  \label{eqn:mqdt-matrix-element-multipole}
\end{align}
where $\hat{Y}_{\kappa q}$ are spherical harmonic operators up to prefactors.
The operators act only on the orbital angular momentum $\lr$ of the Rydberg electron.
Their matrix elements $\braopket{\phi_i}{\hat{Y}_{\kappa q}}{\phi'_j}$ are only non-zero if the quantum numbers of the core are the same and can be evaluated using angular momentum algebra~\cite{edmonds_angular_1957,robicheaux_theory_2018}.
The radial integral only involves the Rydberg electron, since we ignore contributions from core electrons.

Angular momentum operators $\hat{f}_{q}$ do not act on the radial part of the wave function.
Their matrix elements are given by
\begin{align}
  \braopket{\Psi}{\hat{f}_{q}}{\Psi'} = \sum_{i,j} A_i^* A'_j \braopket{\phi_i}{\hat{f}_{q}}{\phi'_j} \int_0^\infty dr \, r^{2} \psi_i(r) \psi'_j(r)
  \,.
  \label{eqn:mqdt-matrix-element-magnetic}
\end{align}
Note that the radial overlap integral does not vanish even if $\nu_i \neq \nu_j$, since the radial wave functions for different $\nu_i$ are not necessarily orthogonal to each other.
The angular matrix elements $\braopket{\phi_i}{\hat{f}_{q}}{\phi'_j}$ can again be evaluated using angular momentum algebra~\cite{edmonds_angular_1957,robicheaux_theory_2018}.

The MQDT formalism thus provides a practical route for computing eigenstates, eigenenergies, and multipole matrix elements of divalent Rydberg atoms.

\section{Green's Tensor Formalism\label{sec:green-tensor}}
In this section, we review the Green's tensor formalism for describing the coherent and dissipative interactions between multilevel atoms mediated by a structured electromagnetic environment.
Relevant realizations include, for example, Rydberg atoms near planar surfaces, atoms trapped close to nanostructures, or atoms in cavities~\cite{jones_modified_2018,dung_resonant_2002,buhmann_dispersion_2012,asenjo-garcia_atomlight_2017,fuchs_casimirpolder_2017,ribeiro_casimirpolder_2015,block_van_2017,block_casimirpolderinduced_2019}.

The electromagnetic environment is fully characterized by the Green's tensor $\tensor{G}(\vec{r},\vec{r}',\omega)$, which describes the propagation of a field of frequency $\omega$ between points $\vec{r}$ and $\vec{r}'$.
The Green's tensor can generally be decomposed into a free space (or bulk) contribution and a scattering contribution as
\begin{align}
  \tensor{G}(\vec{r}, \vec{r}', \omega) = \tensor{G}^0(\vec{r}, \vec{r}', \omega) + \tensor{G}^R(\vec{r}, \vec{r}', \omega)
  \,.
\end{align}
Here, $\tensor{G}^0$ describes propagation in a homogeneous medium and is given by
\begin{align}
  \tensor{G}^0(\vec{r}, \vec{r}', \omega) =
  \left[ \tensor{I} + \frac{1}{k^2} \nabla \nabla \right]
  \frac{e^{\mathrm{i}k|\vec{r} - \vec{r}'|}}{4\pi |\vec{r} - \vec{r}'|}
  \,,
\end{align}
with $k^2 = \varepsilon(\omega)\omega^2/c^2$, where $\varepsilon(\omega)$ is the permittivity (generally complex and frequency-dependent) of the homogeneous medium.
Note that SI units are used throughout this section.
On the other hand, $\tensor{G}^R$ accounts for modifications due to boundaries or material inhomogeneities~\cite{tomas_green_1995,jackson_classical_1998,tomas_recursion_2010}.

A particularly paradigmatic and experimentally relevant class of environments is given by planar layered geometries, which describe, for example, atoms near surfaces or multilayer nanostructures~\cite{buhmann_dispersion_2012,paulus_accurate_2000,tomas_green_1995}.
In such systems, translational invariance parallel to the interfaces allows the Green's tensor to be decomposed into plane-wave components characterized by the in-plane wave vector.
Here, the scattering part $\tensor{G}^R$ accounts for reflections at the (planar) interfaces and depends on the Fresnel reflection coefficients for $s$- and $p$-polarized waves~\cite{jones_modified_2018}.

In the following, we focus on the static limit, where the interatomic and atom-surface distances are much smaller than the wavelength of the relevant transitions.
In this regime, the scattering Green's tensor simplifies considerably and can often be related to electrostatic image-charge solutions.

The interaction between atoms and the electromagnetic field is described by the multipolar coupling Hamiltonian in terms of the electric field operator.
In the following, we restrict the discussion to the electric-dipole approximation for clarity of presentation.
Magnetic multipole contributions are neglected, as they are much weaker for the systems considered here.
With these approximations, the total Hamiltonian consists of three terms:
the bare atomic part $\hat{H}_{0,\alpha}$ for each atom $\alpha$, the bare electromagnetic-field part $\hat{H}_\mathrm{em}$,
and the electric-dipole interaction $\hat{H}_{\mathrm{int},\alpha}$ between each atom $\alpha$ and the field.
This last term gives rise to both dipole-dipole interaction between atoms and self-interaction of each atom in an environment.
The total Hamiltonian then reads~\cite{buhmann_dispersion_2012}
\begin{align}
  \hat{H} &= \sum_{\alpha} \hat{H}_{0,\alpha} + \hat{H}_\mathrm{em} +  \sum_{\alpha} \hat{H}_{\mathrm{int},\alpha}
  \nonumber \\
  &= \sum_{\alpha,n}\hbar \omega_n \ketbra{n_\alpha}{n_\alpha}
  + \!\!\int\!\! \mathrm{d}^3\vec{r}\!\! \int_0^\infty\!\!\!\!\! \mathrm{d}\omega \, \hbar\omega \, \vechat{f}^\dagger(\vec{r}, \omega) \vechat{f}(\vec{r}, \omega)
  \nonumber \\
  &\quad - \sum_{\alpha} \int_0^\infty \mathrm{d}\omega \left[ \vechat{D}^{\alpha} \cdot \vechat{E}(\vec{r}_\alpha, \omega) + \hc \right]
  \,,
  \label{eq:H_total_full}
\end{align}
where $\vechat{f}(\vec{r},\omega)$ are bosonic operators representing the medium-assisted electromagnetic field.
Each atom is modeled as a set of energy eigenstates $\ket{n}$ with energies $\hbar \omega_n$.
The atomic dipole operator is given by
\begin{equation}
  \vechat{D}^{\alpha} = \sum_{n,m} \vec{d}^{\alpha}_{nm} \ketbra{n_\alpha}{m_\alpha}
  \,,
\end{equation}
where $\vec{d}^{\alpha}_{nm}$ are the transition dipole moments.
The electric field operator at the atomic positions can be expressed in terms of the Green's tensor as~\cite{buhmann_dispersion_2012}
\begin{equation*}
  \vechat{E}(\vec{r}_\alpha, \omega) = \mathrm{i} \sqrt{\frac{\hbar}{\pi\varepsilon_0}} \frac{\omega^2}{c^2}
  \!\int \!\mathrm{d}^3 \vec{r} \sqrt{\varepsilon_i(\vec{r},\omega)}  \tensor{G}(\vec{r}_\alpha, \vec{r}, \omega) \, \vechat{f}(\vec{r},\omega)
  \,,
\end{equation*}
where $\varepsilon_0$ is the vacuum permittivity and $\varepsilon_i(\vec{r},\omega)$ is the imaginary part of the medium permittivity, representing absorption.

In the static limit, and after tracing out the field degrees of freedom and applying the Born-Markov approximation, the time evolution of the reduced atomic density matrix $\hat{\rho}$ can be expressed as
\begin{align}
  \frac{\mathrm{d}\hat{\rho}}{\mathrm{d}t}
  &= - \frac{\mathrm{i}}{\hbar} \left[ \sum_{\alpha} \left(\hat{H}_{0,\alpha} + \hat{H}_{\mathrm{si},\alpha}\right) + \hat{H}_{12}, \hat{\rho} \right]
  + \mathcal{L}_\mathrm{diss}[\hat{\rho}]
  \label{eq:master_eq}
  \,.
\end{align}
The term $\mathcal{L}_\mathrm{diss}[\hat{\rho}]$ accounts for dissipative emission of photons into the environment.
We do not consider this term in the following, nor is it included in the software implementation.
Instead, the software calculates Rydberg lifetimes via Einstein coefficients, which is accurate for weak dissipation.

The coherent contributions include the dipole-dipole interaction
\begin{equation}
  \hat{H}_{12}
  = -\frac{1}{2 \varepsilon_0} \sum_{\alpha\neq\beta} \vechat{D}^{\alpha} \tensor{S}(\vec{r}_\alpha, \vec{r}_\beta) \vechat{D}^{\beta}
  \label{eq:H_dd}
\end{equation}
with $\tensor{S}(\vec{r}_\alpha, \vec{r}_\beta) = \lim_{\omega \to 0} \frac{\omega^2}{c^2} \mathrm{Re}\,\tensor{G}(\vec{r}_\alpha, \vec{r}_\beta, \omega)$.
In free space, this expression is equivalent to the dipole-dipole interaction obtained from the analysis of charge distributions that interact via Coulomb forces~\cite{jackson_classical_1998}.

In addition, each atom experiences a self-interaction due to the environment, described by
\begin{equation}
  \hat{H}_{\mathrm{si},\alpha}
  = -\frac{1}{2\varepsilon_0} \vechat{D}^{\alpha} \tensor{S}^R(\vec{r}_\alpha, \vec{r}_\alpha) \vechat{D}^{\alpha}
  \label{eqn:selfinteraction}
\end{equation}
with $\tensor{S}^R(\vec{r}_\alpha, \vec{r}_\beta) = \lim_{\omega \to 0} \frac{\omega^2}{c^2} \mathrm{Re}\,\tensor{G}^R(\vec{r}_\alpha, \vec{r}_\beta, \omega)$.
Here, only the scattering part of the Green's tensor $\tensor{G}^R$ contributes, as the free space contribution is already accounted for in the atomic energies $\omega_n$ through the Lamb shift.
The self-interaction term \cref{eqn:selfinteraction} describes both environment-induced energy shifts of individual atomic levels (Casimir-Polder shifts)~\cite{casimir_influence_1948} and couplings between atomic states~\cite{donaire_casimirpolderinduced_2015}, going beyond what is typically done for Rydberg atoms.
\Cref{sec:application-green-tensor} further illustrates the importance of accounting for both effects.

Finally, we note that no rotating-wave approximation has been applied, such that processes involving off-resonant couplings are retained, which are essential for capturing van der Waals interactions.

\section{Software Implementation\label{sec:implementation}}

\begin{figure}[t]
  \centering
  \includegraphics[width=\columnwidth]{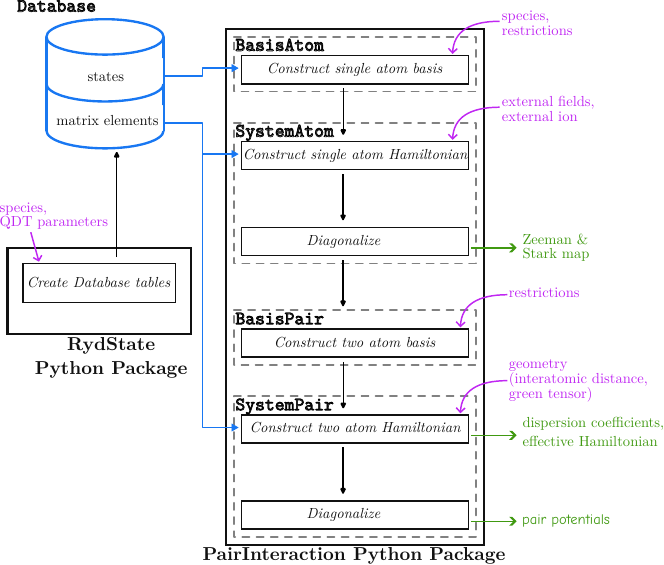}
  \caption{
    \textbf{Schematic overview of the software architecture.} We separate the calculation of states and matrix elements from the generation and analysis of Hamiltonians for one and two Rydberg atoms.
    The former is performed by the RydState Python package.
    The latter is implemented in the PairInteraction Python package, which defines the classes \texttt{BasisAtom}/\texttt{BasisPair} and \texttt{SystemAtom}/\texttt{SystemPair} to represent the objects required to describe the physics.
  }\label{fig:software}
\end{figure}

The discussed Green's tensor formalism and MQDT are implemented in the open-source toolkit PairInteraction v2, which constitutes a complete rewrite of the previous version, PairInteraction v0.9~\cite{weber_calculation_2017}.

The software allows for the construction of Hamiltonians for one or two Rydberg atoms, optionally in the presence of external fields and/or surfaces (see \cref{sec:application-green-tensor}).
The Hamiltonians can be diagonalized to obtain Zeeman and Stark maps (see \cref{sec:application-yb}) or pair-potential curves non-perturbatively.
Moreover, perturbation theory is employed to obtain dispersion coefficients, such as the van der Waals coefficient, and effective Hamiltonians for energetically separated subspaces, which can serve as building blocks for many-body Hamiltonians (see \cref{eq:effective-hamiltonian-spin-1/2,eq:effective-hamiltonian-spin-1} in \cref{sec:application-yb}).
In addition, single-atom properties, such as lifetimes, can be calculated.

In the following, we explain the architecture of the software and provide a conceptual introduction to its usage.
For tutorials and up-to-date software documentation, see our online documentation~\cite{pairinteraction}.

\subsection{Architecture\label{sec:architecture}}

The PairInteraction toolkit is designed to facilitate the construction and diagonalization of Hamiltonians that describe systems consisting of one or two Rydberg atoms in the presence of structured electromagnetic environments, as well as static electric and magnetic fields.
To achieve this efficiently, the calculation of single-atom properties is separated from the Hamiltonian construction by use of databases that contain precomputed Rydberg state eigenenergies, their quantum numbers, and matrix elements between them.
This separation of concerns has multiple benefits: Precomputed database tables avoid the time-consuming computation of Rydberg states and matrix elements.
In addition, this modular architecture allows for the addition of new atomic species and even Rydberg ions by simply adding new database tables.
In principle, any quantum object whose interactions can be described by electric multipole interactions could be incorporated.
The database is generated using the Python package RydState, which implements single- and multi-channel quantum defect theory.
Both the databases and the associated Python package are publicly available online.
The Python package PairInteraction, which enables the construction of Hamiltonians and the analysis of systems comprising one or two Rydberg atoms, automatically downloads the database tables.

The Python API of the PairInteraction package itself follows a domain-modeling approach~\cite{evans_domaindriven_2004}, as shown in \cref{fig:software}, where the objects required to describe the physics are represented by the following classes:
\begin{itemize}
  \item \texttt{BasisAtom}: basis for a Hilbert space of a single Rydberg atom.
    It consists of a list of canonical single-atom states (\texttt{KetAtom} objects), e.g., Rydberg states obtained from MQDT, and a coefficient matrix transforming the canonical states to the basis states.
  \item \texttt{SystemAtom}: system of a single Rydberg atom.
    It consists of a Hamiltonian including external fields expressed as a sparse matrix in a \texttt{BasisAtom} basis.
  \item \texttt{BasisPair}: basis for a Hilbert space of two Rydberg atoms.
    It consists of a list of basis states constructed from pairs of single-atom eigenstates (\texttt{KetPair} objects) and a coefficient matrix transforming the \texttt{KetPair} states to the basis states.
  \item \texttt{SystemPair}: system of two Rydberg atoms.
    It consists of a Hamiltonian including pair interactions expressed as a matrix in a \texttt{BasisPair} basis.
\end{itemize}
The ket objects provide methods for accessing information about Rydberg states such as quantum numbers and lifetimes.
To construct a \texttt{KetAtom} object, a user can specify a Rydberg state by providing approximate quantum numbers of their choice, which the software then uses to find the closest matching canonical single-atom state (with MQDT, most quantum numbers are not good quantum numbers, so the user-specified values are compared to the average values of the quantum numbers).
The \texttt{System} objects provide methods for diagonalizing the Hamiltonians and analyzing them.

To enable the construction and diagonalization of Hamiltonians for large Hilbert space dimensions, computationally expensive calculations are implemented and parallelized in C++, using Intel MKL~\cite{intelmkl} for linear algebra operations and DuckDB~\cite{duckdb} for fast access to precomputed matrix elements and the construction of single-atom operators.
We also provide a wrapped C++ function for diagonalizing multiple Hamiltonians in parallel from Python, significantly improving diagonalization speed compared to the previous version of PairInteraction.
High-level functionality and convenience classes are implemented entirely in Python.
For example, we provide routines to compute Green's tensors for systems of planar surfaces.
This architecture allows new features to be added without modifying the C++ core, lowering the barrier for community-driven extensions.

For frequently recurring tasks, the software additionally provides a graphical user interface (GUI).
The GUI exposes common workflows such as Zeeman and Stark map calculations, pair-potential calculations, and the extraction of dispersion coefficients.
In contrast to the previous version, PairInteraction v0.9, which used a separate GUI, the current GUI is built directly on top of the Python library.
This design makes it straightforward to integrate new library features into the GUI\@.

\subsection{Usage}

In order to illustrate the implementation and facilitate future community contributions, we outline a typical workflow using the Python API and briefly describe the underlying operations performed by the software.
This workflow is applicable to the calculation of, for example, Zeeman and Stark maps for a single-atom target state $\ket{\Psi}$, or pair-potential curves for a two-atom state $\ket{\Psi_1, \Psi_2}$.
In general, $\ket{\Psi_1}$ and $\ket{\Psi_2}$ can belong to different atomic species.

\subsubsection{Constructing a System of a Single Atom}

A basis of the single-atom Hilbert space is created by instantiating the \texttt{BasisAtom} class.
Its constructor allows the basis to be restricted by quantum numbers.
In practice, the principal and orbital quantum numbers, $n$ and $\lr$, can often be constrained to be close to those of the target state $\ket{\Psi}$.
The reasoning is that the target state predominantly couples to states with similar quantum numbers, whereas couplings to states with significantly different values require processes of high order, which are suppressed if states are not energetically degenerate.
Based on the user specifications, states are retrieved from the database.

Using a single-atom basis, one can create an instance of the \texttt{SystemAtom} class.
This class takes as inputs physical entities whose effects can be captured by a single-atom Hamiltonian.
For example, the user may set external electric and magnetic fields or the potential of an ion.
The Hamiltonian is assembled as a matrix in the provided single-atom basis using precomputed matrix elements from the database.

\subsubsection{Constructing a System of Two Atoms}

For pair-potential calculations, two single-atom bases are constructed as described in the previous subsection: one around $\ket{\Psi_1}$ and one around $\ket{\Psi_2}$.
A basis of the two-atom Hilbert space is then created by instantiating the \texttt{BasisPair} class with two diagonalized single-atom systems.
The basis states are direct products of single-atom eigenstates, with energies given by the sum of their eigenenergies.
The energies include all single-atom effects, in particular, Zeeman and Stark shifts.
Using the shifted energy of the target state $\ket{\Psi_1,\Psi_2}$, the basis can be truncated to energetically close pair states.
This basis-construction approach has the advantage that the truncation does not compromise the accuracy of the treatment of single-atom physics.
In addition, the basis can be symmetrized and restricted to states having the same symmetry as the target state~\cite{weber_calculation_2017}.

Using the pair basis, one can create an instance of the \texttt{SystemPair} class.
The user can set the positions of the atoms and, optionally, the position of dielectric plates.
The Hamiltonian, including the interatomic interaction, is then constructed in the provided basis.
To enhance efficiency, the required multipole operators are first constructed for each atom in its single-atom eigenbasis, and the two-atom operators are built from tensor products of these operators.
Crucially, our implementation applies the energy restriction of the basis during the construction of the tensor products.
Thus, Hamiltonian matrix elements outside the selected energy range are never formed.
This substantially reduces computational cost and constitutes a significant improvement over previous software, making previously inaccessible parameter regimes tractable.

\subsubsection{Analyzing a System\label{sec:implementation:analyzing}}
Once a \texttt{SystemAtom} or \texttt{SystemPair} object is created, it can be diagonalized to obtain the eigenenergies and to analyze how these shift as a function of the magnetic and electric fields (Zeeman and Stark maps) or the interatomic distance (pair-potential curves).
To identify the eigenenergies associated with experimentally accessible states, the overlap between eigenstates and target states can be calculated.

In addition, the PairInteraction toolkit provides functionality to compute dispersion coefficients, such as van der Waals coefficients, and effective Hamiltonians using perturbation theory.
While calculations are internally performed in atomic units, integration with the Pint unit library~\cite{pint} allows all quantities to be specified in arbitrary units, facilitating direct comparison with experiments.

\subsection{Benchmarks\label{sec:benchmarks}}
\begin{figure}
  \centering
  \includegraphics{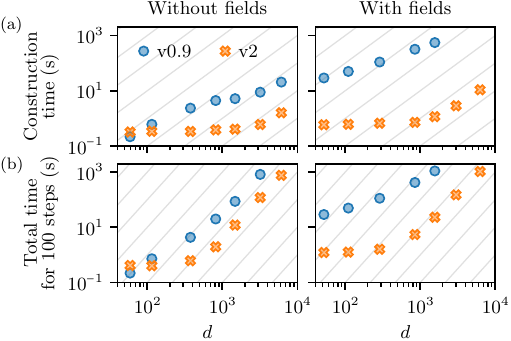}
  \caption{\textbf{Comparison of the computational performance of PairInteraction v2 with v0.9} (on an AMD Ryzen 7 5700G with \qty{64}{GiB} of RAM).
    A system of two interacting rubidium atoms is considered, with and without electric and magnetic fields of strength $E_z=\qty{0.2}{V/cm}$ and $B_z=\qty{100}{G}$.
    We measure the wall-clock time for (a) constructing a two-atom Hamiltonian and (b) computing interaction potentials for 100 interatomic distances ranging from $2$ to \qty{3}{\mu m} for the target state $\state{63P_{1/2},m=1/2;\,63P_{1/2},m=1/2}$.
    The benchmarks assume all database tables have been downloaded before.
    The Hamiltonian is constructed in a basis restricted to symmetrized pair states~\protect{\cite{weber_calculation_2017}} with $n=59\ldots67$ and $l=0\ldots5$, whose energies differ at most by $\Delta E$ from the target state.
    $\Delta E$ is varied between $2$ and \qty{128}{GHz} to run the benchmarks for two-atom Hamiltonians of different dimensions $d$.
    The single-atom Hilbert space dimension is $d_1=648$ for all runs.
    The gray lines indicate the asymptotic scaling with $d^2$ for the construction time and $d^3$ for the total time.
  }\label{fig:benchmark}
\end{figure}

We benchmark the new PairInteraction version v2 against the previous version v0.9.

We observe speedups of one order of magnitude in the construction of Hamiltonians for two interacting Rydberg atoms and in the calculation of interaction potentials, see \cref{fig:benchmark}.
In the presence of static electric and magnetic fields, the speedups can even reach nearly two orders of magnitude.
The speedups result from parallelizing the diagonalization over interatomic distances in C++, using Intel MKL for linear algebra operations and DuckDB for constructing single-atom operators from precomputed matrix elements, and from computing the two-atom Hamiltonian via energy-restricted tensor products in single-atom eigenbases, as described in the previous sections.
In addition, we gained a factor of two by allowing the use of single-precision floating-point numbers instead of double precision for diagonalization.
The resulting eigenenergies deviate from double-precision results by only about $10^{-7}$ relative to the Hamiltonian's spectral width, with absolute deviations typically below \qty{10}{kHz}.

To understand the runtime for large systems, we vary the two-atom Hilbert-space dimension $d$ while keeping the single-atom dimension fixed at $d_1$.
Because the two-atom basis is energy-truncated, it can be $d<d_1$.
In this case, the runtime is dominated by constructing and diagonalizing single-atom operators.
For large $d$, the runtime follows expected scaling laws, indicated by the gray guidelines in \cref{fig:benchmark}: The construction time approximately scales as $\mathcal{O}(d^2)$, reflecting the number of entries in the two-atom Hamiltonian.
The total time for calculating pair potentials scales as $\mathcal{O}(d^3)$, dominated by the diagonalization.
Note that caching is applied to suppress the relative cost of Hamiltonian construction when many interatomic distances are evaluated.

\section{Example Applications\label{sec:applications}}

We illustrate the capabilities of the PairInteraction toolkit with two example applications.
The code for these examples is available in the PairInteraction online documentation~\cite{pairinteraction}.

In \cref{sec:application-yb}, we perform calculations for $^{174}$Yb, an atomic species with two valence electrons that requires the MQDT formalism of \cref{sec:mqdt}.
By comparing calculated and experimentally measured Stark maps near an avoided crossing, we demonstrate that energy levels and dipole matrix elements are computed accurately, where we use MQDT models from Refs.~\cite{peper_spectroscopy_2025,kuroda_microwave_2025}.
We then extend our analysis to two atoms, study pair interaction potentials for $^{174}$Yb, and show how to construct effective spin-$1/2$ and spin-$1$ Hamiltonians using the PairInteraction toolkit.

In \cref{sec:application-green-tensor}, we use the Green's tensor formalism of \cref{sec:green-tensor} to study the self-interaction as well as the Rydberg-Rydberg interactions near a perfectly conducting plate.

\subsection{Modeling One and Two \texorpdfstring{$^{174}$}{174}Yb Atoms\label{sec:application-yb}}
\begin{figure}
  \centering
  \includegraphics{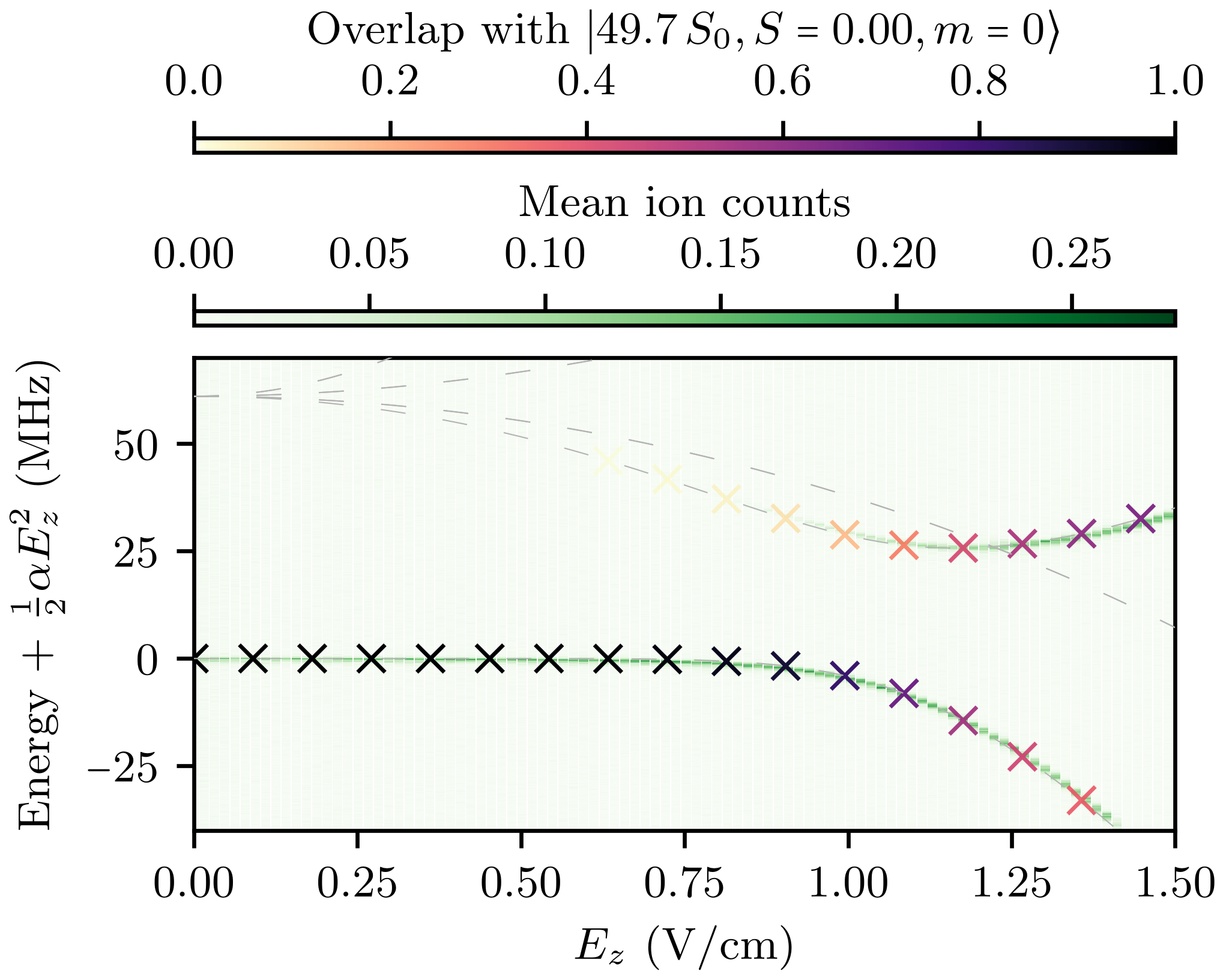}
  \caption{\textbf{Comparison of a measured and calculated Stark map in $^{174}$Yb.}
    The Stark map shows an avoided crossing of the $\state{49.7\, S_0, \stot=0.00, m=0}$ with the nearby $\state{49.7\, F_3, \stot=0.00, m=0}$ state.
    The green colormap shows the ion counts in the experiment for the given electric field and two-photon detuning, plotted as energy on the $y$-axis.
    The gray lines are the calculated energy levels, and the colored markers additionally show the overlap of the eigenstate with the state of interest.
    All energies are shifted by a quadratic Stark shift $\frac{1}{2} \alpha E_z^2$ with static polarizability $\alpha$ extracted from the low-field regime.
  }\label{fig:starkmapYb174}
\end{figure}
As discussed in \cref{sec:mqdt-labeling}, for divalent atoms such as $^{174}$Yb, not all quantum numbers are necessarily good quantum numbers.
In the following, the states of interest are best described by the LS-coupling scheme, where the total orbital angular momentum $\ltot$ is still an almost good quantum number, but the total spin $\stot$ can mix.
Thus, we label the states as $\state{\nu\, {\ltot}_{\jtot}, \stot, m}$, stating the averaged total spin $\stot$ explicitly.

To benchmark the implementation of PairInteraction, we compare a calculated Stark map for $^{174}$Yb with an experimentally measured one, see \cref{fig:starkmapYb174}.
The measurement was performed in a cold atomic cloud by exciting to $\state{49.7\, S_0, \stot=0.00, m=0}$ via a two-photon scheme, and detecting the resulting Rydberg states via state-selective electric-field ionization~\cite{wang_twocolor_2025}.
To match the experimental results to the calculated Stark map, we rescaled the electric-field values for the experimental data by a factor of $\num{1.028}$, which is well within the tolerances of the simulations used to calculate the field strength within the experimental apparatus.
Notably, Stark maps recorded for different Rydberg states are all brought into agreement with theory using the same scaling factor.
Interestingly, $^{174}$Yb has an accidental near-degeneracy of the $S_0$ and $F_3$ Rydberg series.
Although the electric field couples these states only in third order, the small energy separation strongly enhances the mixing at relatively low fields of $\sim \qty{1}{V/cm}$, giving the $F_3$ enough $S_0$ character to be addressable by the two-photon excitation scheme.
We achieve excellent agreement between theory and experiment for the avoided crossing of these states.
Because of the involved higher-order process, this not only validates the employed MQDT models but also demonstrates the accuracy of calculated energies and matrix elements for multiple states.

\begin{figure}
  \centering
  \includegraphics{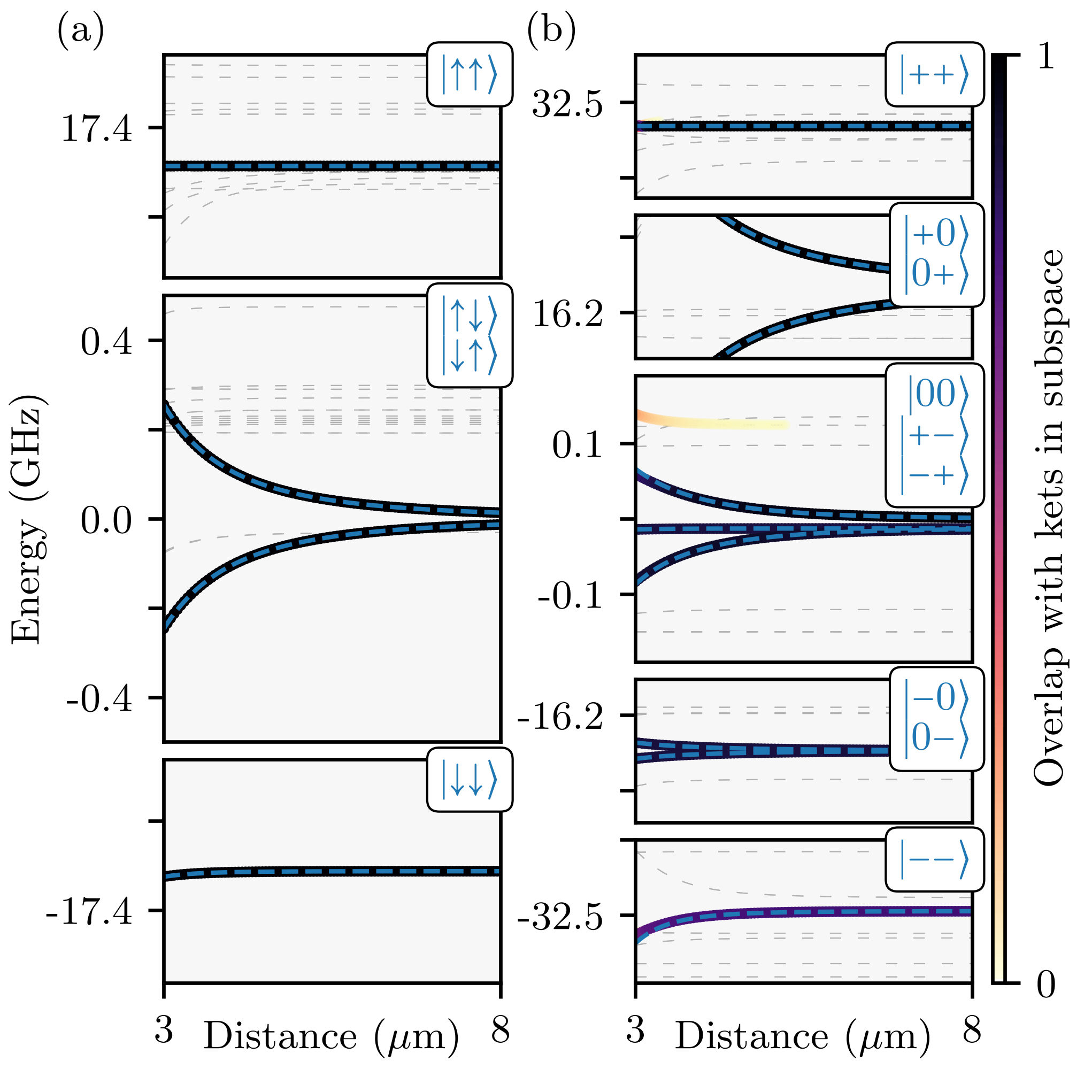}
  \caption{\textbf{Pair-interaction potentials for two $^{174}$Yb atoms as a function of internuclear distance.}
    The internuclear axis is chosen along the quantization axis $z$.
    (a) At a magnetic field of $B_z=\qty{20}{G}$ for all pair states formed from $\ket{\downarrow}$ and $\ket{\uparrow}$, as defined in \cref{eq:spin-1/2-states}, spanning an effective spin-$1/2$ subspace.
    (b) At a magnetic field of $B_z=\qty{350}{G}$ for all pair states formed from $\ket{-}$, $\ket{0}$, and $\ket{+}$, as defined in \cref{eq:spin-1-states}, spanning an effective spin-$1$ subspace.
    In gray, the exact pair potentials calculated in the full basis are shown.
    The overlap with the states in the target subspace is indicated by the colored points.
    The blue dotted lines show the pair-potential curves obtained from the effective Hamiltonian, which is constructed via perturbation theory up to second order.
  }\label{fig:effective_spin_hamiltonian}
\end{figure}
We continue by analyzing pair interactions for two $^{174}$Yb atoms and show how these can be used to construct effective spin Hamiltonians.
Let us start by looking at the two states
\begin{align}
  \ket{\uparrow} &= \state{50.1\, P_1, \stot=0.05, m=0}
  \text{ and } \nonumber \\
  \ket{\downarrow} &= \state{49.7\, S_0, \stot=0.00, m=0}
  \,,
  \label{eq:spin-1/2-states}
\end{align}
which we label as spin-$1/2$ states.
Their pair interaction potentials are shown in \cref{fig:effective_spin_hamiltonian}(a).
The state $\ket{\downarrow\downarrow}$ on the bottom has a small attractive $V_6 = \abs{C_6}/r^6$ van der Waals potential, whereas the states $\ket{\uparrow\downarrow}$ and $\ket{\downarrow\uparrow}$ are coupled via resonant $V_3 = C_3/r^3$ dipole-dipole interaction.

In general, the total Hamiltonian for an effective spin-$1/2$ model based on the states in \cref{eq:spin-1/2-states} can be split into a single-particle term
$\hat{H}_\alpha = E_\downarrow \ketbra{\downarrow}{\downarrow} + E_\uparrow \ketbra{\uparrow}{\uparrow}$,
and an interaction term
\begin{equation}
  \hat{H}_\mathrm{int}
  = \begin{pmatrix}
    V_6^{\uparrow\uparrow} & 0                        & 0                        & 0                          \\
    0                      & V_6^{\uparrow\downarrow} & V_3                      & 0                          \\
    0                      & V_3                      & V_6^{\uparrow\downarrow} & 0                          \\
    0                      & 0                        & 0                        & V_6^{\downarrow\downarrow}
  \end{pmatrix}
  \,.
  \label{eq:effective-hamiltonian-spin-1/2}
\end{equation}
The interaction Hamiltonian has a block-diagonal structure, where the blocks are defined by the energetically separated subspaces of the pair states, analogously to the subspaces in \cref{fig:effective_spin_hamiltonian}(a).
If one then studies many-body systems, one should also check that no three-atom resonances appear with unwanted states, and the Hilbert space is still well described by the Rydberg states of interest.
Such effective Hamiltonians have already been widely used to study spin-$1/2$ models in Rydberg systems~\cite{barredo_coherent_2015,deleseleuc_observation_2019,browaeys_manybody_2020,morgado_quantum_2021,lippe_experimental_2021,chen_continuous_2023,franz_observation_2024}.

Instead of the two-level structure above, one might want to achieve a three-level structure, e.g., to model a spin-$1$ system.
This can be obtained by tuning, for example, the $\ket{\downarrow\downarrow}$ state into a Förster resonance with another pair state by changing the principal quantum number and using a magnetic field.
In this example, the states of interest are given by
\begin{align}
  \ket{+} &= \state{51.1\, P_1, \stot=0.04, m=0}
  \,, \nonumber \\
  \ket{0} &= \state{50.7\, S_0, \stot=0.00, m=0}
  \,,\text{ and } \nonumber \\
  \ket{-} &= \state{50.4\, P_1, \stot=0.95, m=0}
  \,,
  \label{eq:spin-1-states}
\end{align}
where the states are labeled as spin-$1$ states.
For a magnetic field of \qty{350}{G}, the states are energetically close to a Förster resonance, i.e., the energy difference of the states $\ket{0, 0}$ and $\ket{-, +}$ is on the same order as their interaction strength.
Their pair interaction potentials are shown in \cref{fig:effective_spin_hamiltonian}(b).

Without loss of generality, we can choose $E_0 = D$ and $E_\pm = \pm \omega$, where $D$ is on the order of the interaction strength and $\omega$ is large compared to both.
Then, the single-atom Hamiltonian is given by
$\hat{H}_\alpha = - \omega \ketbra{-}{-} + D \ketbra{0}{0} + \omega \ketbra{+}{+}$.

The two-atom interaction Hamiltonian is a $9\times 9$ matrix, which, similarly to \cref{fig:effective_spin_hamiltonian}(b), can be split into five energetically separated subspaces.
The complete interaction Hamiltonian for a three-level system such as that defined in \cref{eq:spin-1-states} in general is given by
\begin{equation}
  \scalebox{0.9}{$\displaystyle
  \hat{H}_\mathrm{int}
  = \begin{pmatrix}
    V_6^{++} &          &          &          &          &          &          &          &          \\
        & V_6^{+0} & V_3^{+0} &          &          &          &          &          &          \\
        & V_3^{+0} & V_6^{+0} &          &          &          &          &          &          \\
        &          &          & V_6^{-+} & V_3      & B        &          &          &          \\
        &          &          & V_3      & V_6^{00} & V_3      &          &          &          \\
        &          &          & B        & V_3      & V_6^{-+} &          &          &          \\
        &          &          &          &          &          & V_6^{-0} & V_3^{-0} &          \\
        &          &          &          &          &          & V_3^{-0} & V_6^{-0} &          \\
        &          &          &          &          &          &          &          & V_6^{--} \\
  \end{pmatrix}
  $}
  \,.
  \label{eq:effective-hamiltonian-spin-1}
\end{equation}
Here, $V_3$ describes the dipole-dipole coupling of the state $\ket{00}$ to the states $\ket{+-}$ and $\ket{-+}$, and $B \propto r^{-6}$ describes potential second-order couplings between the states $\ket{-+} \leftrightarrow \ket{+-}$.
This Hamiltonian structure can be used to describe spin-$1$ systems, where spin conservation is again enforced by the energy separation of the subspaces~\cite{mogerle_spin1_2025,liu_supersolidity_2024,qiao_realization_2025}.

Constructing such effective Hamiltonians is straightforward using the PairInteraction toolkit.

\subsection{Rydberg Atoms near a Surface\label{sec:application-green-tensor}}
\begin{figure}
  \centering
  \includegraphics{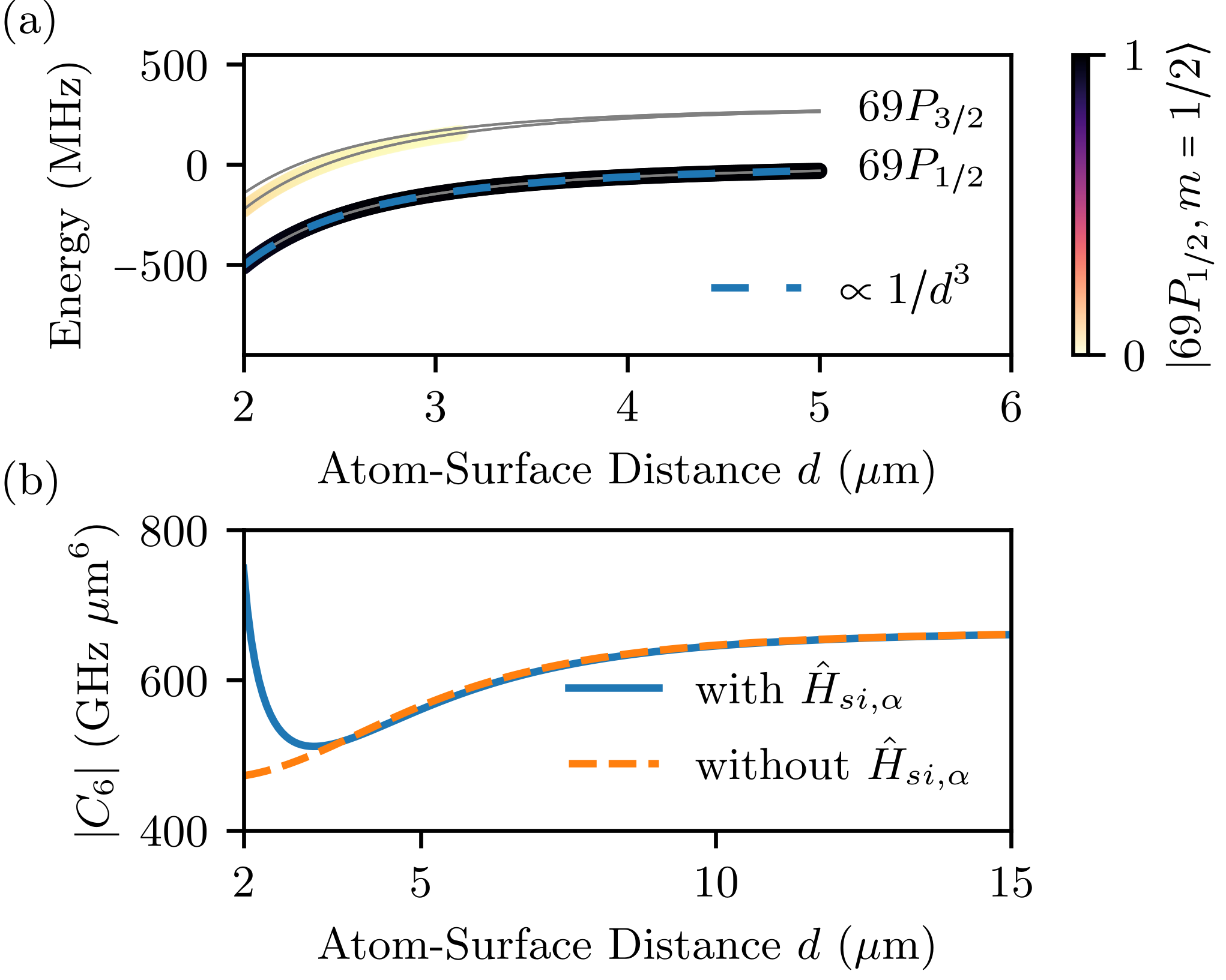}\\
  \caption{\textbf{Rydberg interactions near a perfectly conducting plate.}
    (a) The self-interaction of a single Rydberg atom near a plate at distance $d$.
    (b) The $C_6$ coefficient of the pair state $\state{69 S_{1/2}, m=1/2; 72 S_{1/2}, m=1/2}$ of two rubidium atoms as a function of the distance $d$ to the plate.
    For comparison, we performed the calculation with (blue solid line) and without (orange dashed line) including the self-interaction.
    The two atoms are aligned along $z$ at a distance of $r=\qty{10}{\mu m}$.
    Both atoms have the same distance $d$ to the plate, whose normal vector points along $x$.
  }\label{fig:green_tensor}
\end{figure}
Now we study the effect of a perfectly conducting plate on Rydberg atoms using the Green's tensor formalism described in \cref{sec:green-tensor}.
In \cref{fig:green_tensor}(a), we calculate the self-interaction of a single Rydberg atom near a plate.
The plot shows the Casimir-Polder energy shift of the rubidium Rydberg state $\state{69 P_{1/2}, m=1/2}$, which here in the static limit scales with $1/d^3$.
In addition, for short distances, we observe a level mixing induced by the plate between the $\state{69 P_{1/2}, m=1/2}$ and $\state{69 P_{3/2}, m=1/2}$ states.
For more than one atom, these shifts and level mixings can affect the efficiency of the Rydberg blockade mechanism.

Furthermore, in \cref{fig:green_tensor}(b), we analyze the $C_6$ coefficient of a rubidium pair state $\state{69 S_{1/2}, m=1/2; 72 S_{1/2}, m=1/2}$.
The atoms are placed at a distance $r=\qty{10}{\mu m}$ from each other, and both are at the same distance $d$ from the plate.
For large atom-surface distances $d$, the $C_6$ coefficient approaches its free-space value.
Without self-interactions, we reproduce the behavior of the $C_6$ coefficient previously obtained using an earlier Green's tensor implementation in PairInteraction~\cite{block_van_2017}.
Including the self-interaction, we see clear deviations at small distances, where the self-interaction becomes relevant.
This can be explained by the fact that the $\state{69 P_{3/2}; 72 P_{3/2}}$ manifold, which is primarily responsible for the $C_6$ coefficient, experiences a different self-interaction near the surface and therefore a different energy shift than the pair state $\state{69 S_{1/2}, m=1/2; 72 S_{1/2}, m=1/2}$.

\section{Conclusion and Outlook\label{sec:conclusion}}

We presented a framework for modeling Rydberg atoms and their interactions.
We first reviewed multi-channel quantum defect theory (MQDT) for a precise description of Rydberg states.
Treating Rydberg states as superpositions of multiple channel functions is required if excitations of the ionic core are relevant, as is commonly the case for multivalent atoms such as strontium and ytterbium.
We proposed an approach to consistently label these states by averaged angular quantum numbers.
To model the interaction between Rydberg atoms both in free space and in structured electromagnetic environments, we reviewed the Green's tensor formalism.
This formalism has the benefit of allowing the calculation of interactions in arbitrary geometries, including modifications induced by nearby surfaces.
We presented the key equations of the formalism for multilevel atoms in the static limit.
We discussed self-interactions, including couplings between different states, and showed that, contrary to common approximations in the literature, these couplings cannot generally be neglected for Rydberg atoms near surfaces.
With this additional contribution included, we obtained a non-perturbative, complete treatment of interactions within the static Green's tensor formalism.

We discussed the implementation of the framework in the open-source toolkit PairInteraction v2.
In addition to the added capabilities, we demonstrated that the software is one order of magnitude faster for the calculation of Rydberg pair potentials than its predecessor.
The speedup was achieved by using high-performance libraries for the parallelized diagonalization of Hamiltonians and for caching matrix elements, as well as algorithmic improvements in the construction of Hamiltonians.
We demonstrated example applications, including the comparison of a computed and experimentally measured Stark map of $^{174}$Yb, showing the high accuracy of the MQDT calculations.
The software is available at \url{www.pairinteraction.org}.

A natural direction for future work is to further refine the MQDT description, especially for low-lying states, by using model potentials that more accurately resemble the potential of the ionic core.
In addition, it would be interesting to go beyond the static limit of the Green's tensor to describe, for example, couplings to resonant cavity modes.
Beyond neutral Rydberg atoms, the modular design of the software allows for extending it to other systems, such as dipolar molecules or Rydberg ions;
for example, support for $^{88}\text{Sr}^{+}$ has already been added to PairInteraction.

% % % % % % % % % % ACKNOWLEDGMENTS % % % % % % % % % %
\begin{acknowledgments}
  We would like to thank Florian Meinert, Ming Li, Masaya Kunimi, Toshi Kusano, Pedro Ildefonso, and Aleksei Konovalov for insightful discussions.
  We thank Patrick Mischke and Simon Hollerith for bug testing and contributions to the recent version of PairInteraction.
  We thank Florian Pausewang, Xin Wang, and Eduardo Uruñuela for their work on the measurement, analysis, and interpretation of the Stark map, and for valuable discussions.
\end{acknowledgments}

% % % % % % % % % % FUNDING % % % % % % % % % %
\section*{Funding}
J.M., H.P.B., and S.W.\ acknowledge funding from the Federal Ministry of Research, Technology and Space (BMFTR) under the grants MUNIQC-Atoms and BeRyQC, and from the Deutsche Forschungsgemeinschaft (DFG) under the Priority Programme SPP 2514 and the JST-DFG ASPIRE 2025 project ``Quantum computing with neutral atoms''.
T.L.\ and S.H.\ acknowledge support by the European Union's Horizon 2020 program under the ERC grant SUPERWAVE (grant No.101071882) and by the Deutsche Forschungsgemeinschaft (DFG) within the collaborative research center SFB/TR185 OSCAR, project A8 (No. 277625399).
B.O.\ and A.K.\ acknowledge support by the Deutsche Forschungsgemeinschaft (DFG, German Research Foundation) under the Research Unit FOR 5413/1, Grant No. 465199066, and the IQST Boosting Programme 2026.
E.J.B.\ acknowledges support by the IMPRS for Quantum Dynamics in Physics, Chemistry and Biology.

% % % % % % % % % % AUTHOR CONTRIBUTIONS % % % % % % % % % %
\section*{Author Contributions}
J.M., F.H., H.M., S.W. developed the software, with contributions from A.K., E.J.B.; J.M., F.H., T.L., S.W. worked on MQDT\@; J.M., A.K., B.O. worked on the Green's tensor formalism; J.K., B.O., S.H., H.P.B., S.W. supervised the work; S.W. coordinated the project; all authors contributed to the writing of the manuscript.

% % % % % % % % % % DATA AVAILABILITY % % % % % % % % % %
\section*{Data Availability}
The data generated and analyzed during the current study is openly available~\cite{DARUS-6106_2026}.
\nocite{DARUS-6106_2026}

% % % % % % % % % % BIBLIOGRAPHY % % % % % % % % % %
\bibliography{manuscript} % using bibtex
\end{document}